# Quantum weak invariants: Dynamical evolution of fluctuations and correlations

**Zeyi Shi**[1] **and Sumiyoshi Abe**[1,2,3,4,*]

[1] Department of Physics, College of Information Science and Engineering, Huaqiao University, Xiamen 361021, China

[2] Institute of Physics, Kazan Federal University, Kazan 420008, Russia

[3] Department of Natural and Mathematical Sciences, Turin Polytechnic University in Tashkent, Tashkent 100095, Uzbekistan

[4] ESIEA, 9 Rue Vesale, Paris 75005, France

* Correspondence: suabe@sf6.so-net.ne.jp

**Abstract:** Weak invariants are time-dependent observables with conserved expectation values. Their fluctuations, however, do not remain constant in time. On the assumption that time evolution of the state of an open quantum system is given in terms of a completely positive map, the fluctuations monotonically grow even if the map is not unital, in contrast to the fact that monotonic increases of both the von Neumann entropy and Rényi entropy require the map to be unital. In this way, the weak invariants describe temporal asymmetry in a manner different from the entropies. A formula is presented for time evolution of the covariance matrix associated with the weak invariants in the case where the system density matrix obeys the Gorini-Kossakowski-Lindblad-Sudarshan equation.

## 1. Introduction

Ever since the establishment of the second law of thermodynamics, its relation to time-reversal invariance in the microscopic laws has been remaining as a most outstanding problem in physics [1,2]. It is generally assumed that Eddington is the first who has coined the term *time's arrow* in the context of physics [3]. He mentions by personification: "Shuffling is the only thing which Nature cannot undo". According to him, a *primary law* is a law which shows that it is impossible for some things to occur, whereas a *secondary law* is a law which tells that it is too improbable for them to occur. He also mentions: "The question whether the second law of thermodynamics and other statistical laws are mathematical deductions from the primary laws, presenting their results in a conveniently usable form, is difficult to answer; but I think it is generally considered that there is an unbridgeable hiatus. At the bottom of all the questions settled by secondary law there is an elusive conception of '*a priori* probability of states of the world' which involves an essentially different attitude to knowledge from that presupposed in the construction of the scheme of primary law."

Clearly, a key physical quantity relevant here is the thermodynamic entropy. The present authors interpret that the purpose of this Special Issue is to develop discussions about the relationship between the thermodynamic entropy and time's arrow. It is noted however that such discussions have recently been criticized in a paper published in this journal [4]. There, the second law is strictly limited to the thermodynamics of equilibrium states of matters and any possible generalization of the entropy for time-dependent processes in nonequilibrium thermodynamics is ruled out. This point will not be discussed here, and instead the concept of weak invariants introduced in Reference [5] will be considered as an indicator of asymmetry of time.

Thus, the purpose of the present paper is to study a novel aspect of temporally monotonic behaviors of open quantum systems (i.e., systems surrounded by the environment) by use of weak invariants and to compare it with time evolution of the Rényi entropy as well as the von Neumann entropy. First, a linear and completely positive map describing subdynamics of an open quantum system is recapitulated. Next, the fluctuation of a weak invariant is shown to never decrease under such a map, in contrast to the fact that the entropies exhibit such a monotonic behavior under an additional requirement that the map is unital. Then, a specific case is examined when time evolution is further assumed to be Markovian. Accordingly, what is relevant there is the Gorini-Kossakowski-Lindblad-Sudarshan equation [6-8]. Since there may exist multiple weak invariants, the covariance matrix associated with them is naturally defined. Presented is a formula for time evolution of the covariance matrix under the Gorini-Kossakowski-Lindblad-Sudarshan equation, which generalizes the result given in Reference [9]. A temporally monotonic behavior is observed in the diagonal elements of the covariance matrix. In this way, the fluctuations of the weak invariants are seen to exhibit asymmetry in time.

Throughout this paper, $\hbar$ is set equal to unity.

## 2. Behaviors of Weak Invariant and Entropy under Completely Positive Map

Let us start our discussion with a completely positive map. The total isolated system consists of the objective system $A$ and the environment $B$, which are interacting with each other. Suppose $A$ and $B$ to be unentangled at certain time $t$. Hence, the quantum state of the total system is written as $|\Psi(t)\rangle_{AB} = |\psi(t)\rangle_A \otimes |\phi(t)\rangle_B$. Although it is also possible to consider the state given by the tensor product of the density matrices describing mixed states of $A$ and $B$, the result obtained is similar [10,11]. Since the total system is isolated, the state evolves in time as follows: $|\Psi(t)\rangle_{AB} \to |\Psi(t')\rangle_{AB} = U_{AB}(t',t)|\Psi(t)\rangle_{AB}$ $(t' > t)$, where $U_{AB}$ is the unitary time-evolution operator that is nonlocal (i.e., $U_{AB} \neq U_A \otimes U_B$) because of the interaction between $A$ and $B$. If the state of $B$ is ignored at $t'$, then the reduced density matrix of $A$ is obtained via the partial trace of $|\Psi(t')\rangle_{AB\ AB}\langle\Psi(t')|$ over $B$, that is,

$\rho_A(t') = \mathrm{tr}_B\left(|\Psi(t')\rangle_{AB\ AB}\langle\Psi(t')|\right) = \mathrm{tr}_B\left(U_{AB}(t',t)|\Psi(t)\rangle_{AB\ AB}\langle\Psi(t)|U^\dagger_{AB}(t',t)\right)$. The partial trace may be performed in terms of a certain complete orthonormal system $\{|v_n\rangle_B\}_n$:

$\rho_A(t') = \sum_n {}_B\langle v_n|U_{AB}(t',t)|\phi(t)\rangle_B\, \rho_A(t)\, {}_B\langle\phi(t)|U^\dagger_{AB}(t',t)|v_n\rangle_B$, where $\rho_A(t)$ is a pure state $\rho_A(t) = |\psi(t)\rangle_{A\ A}\langle\psi(t)|$. Therefore, time evolution of the state of the subsystem $A$ is described by a completely positive map

$$\Phi_{t',t}\left(\rho_A(t)\right) \equiv \rho_A(t') = \sum_n V_{A,n}(t',t)\, \rho_A(t)\, V^\dagger_{A,n}(t',t) \qquad (1)$$

in the Kraus representation [10,11] with

$$V_{A,n}(t',t) = {}_B\langle v_n|U_{AB}(t',t)|\phi(t)\rangle_B, \qquad (2)$$

which are operators in the space of $A$. Since the total time-evolution operator is unitary,

$$\sum_n V_{A,n}^\dagger(t',t) V_{A,n}(t',t) = \mathbb{I}_A \qquad (3)$$

holds, where $\mathbb{I}_A$ is the identity operator in the space of *A*. Equation (3) is the trace-preserving condition, $\text{tr}_A \rho_A(t') = \text{tr}_A\left(\Phi_{t',t}\left(\rho_A(t)\right)\right) = \text{tr}_A \rho_A(t) (\equiv 1)$, ensuring the conservation of probability. Therefore, the set $\left\{V_{A,n}^\dagger(t',t) V_{A,n}(t',t)\right\}_n$ forms a positive operator-valued measure commonly abbreviated as POVM in the literature.

If the procedure mentioned above is viewed in the opposite way, then it will be seen as an example of the Naimark extension [12], which allows to express any POVM as the reduction of a projection operators in an extended space (i.e., the total space of *A* and *B* in the present case). This is seen as follows. Taking Equation (2) into account, write $\left|\Psi(t')\right\rangle_{AB} = U_{AB}(t',t)\left|\psi(t)\right\rangle_A \otimes \left|\phi(t)\right\rangle_B = \sum_n \left(V_{A,n}\left|\psi(t)\right\rangle_A\right) \otimes \left|v_n\right\rangle_B$, from which Equation (1) straightforwardly follows, and Equation (3) is nothing but the normalization condition on $\left|\Psi(t')\right\rangle_{AB}$. Define the projection operator: $\Pi_{AB,n}(t',t) = U_{AB}^\dagger(t',t)\left(\mathbb{I}_A \otimes \left|v_n\right\rangle_{BB}\left\langle v_n\right|\right)U_{AB}(t',t)$ satisfying $\Pi_{AB,n}(t',t)\Pi_{AB,n'}(t',t) = \Pi_{AB,n}(t',t) \times \delta_{nn'}$ and $\sum_n \Pi_{AB,n}(t',t) = \mathbb{I}_{AB}$. The probability associated with the "projective measurement" in the extended space, $\text{tr}_{AB}\left(\Pi_{AB,n}(t',t)\left|\Psi(t)\right\rangle_{AB\,AB}\left\langle\Psi(t)\right|\right)$, is calculated to be $\text{tr}_A\left(V_{A,n}(t',t)\rho_A(t)V_{A,n}^\dagger(t',t)\right)$, which is the probability associated with the "POVM measurement".

It is noted that, as long as concerning the map in Equation (1), $\rho_A(t)$ does not have to be a pure state. In fact, it can be an arbitrary mixed state in the subsequent discussion.

A special class of completely positive maps is characterized in the case when the identity operator is a fixed point of the maps. For the map in Equation (1), the fixed point condition, $\Phi_{t',t}(\mathbb{I}_A) = \mathbb{I}_A$, reads

$$\sum_n V_{A,n}(t',t) V_{A,n}^\dagger(t',t) = \mathbb{I}_A. \qquad (4)$$

A map simultaneously satisfying Equations (3) and (4) is called *unital* [13]. Clearly, this condition is fulfilled if $V_{A,n}$'s are *normal* [13], that is, $\left[V_{A,n}(t',t), V_{A,n}^\dagger(t',t)\right] = 0$, which may not hold, in general.

It is instructive to see the effects of the dynamical map $\Phi_{t',t}$ in Equation (1) on the entropic quantities. Consider the von Neumann entropy

$$S[\rho] = -\mathrm{tr}(\rho \ln \rho). \qquad (5)$$

Here and hereafter, the subscript *A* is abbreviated for the sake of notational convenience. The operator-valued function inside this trace operation

$$f(\rho) \equiv -\rho \ln \rho \qquad (6)$$

is operator concave [14], provided that an operator-valued function *f* is said to be concave (convex) if it satisfies $f(\lambda \rho_1 + (1-\lambda)\rho_2) \geq (\leq) \lambda f(\rho_1) + (1-\lambda) f(\rho_2)$ for $0 \leq \lambda \leq 1$, and the operator inequality $\sigma_1 \geq \sigma_2$ implies that both $\sigma_1$ and $\sigma_2$ are Hermitian and all eigenvalues of $\sigma_1 - \sigma_2$ are nonnegative. The concavity of $f(\rho)$ leads to [11]

$$f(\rho(t')) = f(\Phi_{t',t}(\rho(t))) \geq \Phi_{t',t}(f(\rho(t))), \qquad (7)$$

*if the dynamical map* $\Phi_{t',t}$ *is unital*. Therefore, from the trace-preserving nature of the map, the von Neumann entropy is seen to behave as

$$S[\rho(t')] \geq S[\rho(t)] \qquad (8)$$

for $t' > t$. It is possible to generalize this result to the case of the Rényi entropy indexed by positive $\alpha$ ($\alpha \neq 1$) [15]

$$S_\alpha[\rho] = \frac{1}{1-\alpha} \ln(\mathrm{tr}\,\rho^\alpha). \qquad (9)$$

This quantity converges to the von Neumann entropy in the limit $\alpha \to 1$. $g(\rho) \equiv \rho^\alpha$ is operator concave (convex) if $0 < \alpha \leq 1$ ($1 \leq \alpha \leq 2$) [13,14]. (It is not convex if $\alpha > 2$.) Therefore, if the dynamical map $\Phi_{t',t}$ is unital, then holds the following operator inequalities: $\{\Phi_{t',t}(\rho(t))\}^\alpha \geq \Phi_{t',t}(\rho^\alpha(t))$ ($0 < \alpha \leq 1$) and $\{\Phi_{t',t}(\rho(t))\}^\alpha \leq \Phi_{t',t}(\rho^\alpha(t))$ ($1 \leq \alpha \leq 2$), implying that $\rho^\alpha(t') \geq \Phi_{t',t}(\rho^\alpha(t))$ ($0 < \alpha \leq 1$) and $\rho^\alpha(t') \leq \Phi_{t',t}(\rho^\alpha(t))$ ($1 \leq \alpha \leq 2$), and thus $\mathrm{tr}(\rho^\alpha(t')) \geq \mathrm{tr}(\Phi_{t',t}(\rho^\alpha(t))) = \mathrm{tr}(\rho^\alpha(t))$ ($0 < \alpha \leq 1$) and $\mathrm{tr}(\rho^\alpha(t')) \leq \mathrm{tr}(\rho^\alpha(t))$ ($1 \leq \alpha \leq 2$), respectively. Therefore, as the von Neumann entropy, the Rényi entropy does not decrease in time under the unital dynamical map, either:

$$S_\alpha[\rho(t')] \geq S_\alpha[\rho(t)] \qquad (10)$$

for $t' > t$ if $0 < \alpha \leq 2$ ($\alpha \neq 1$). This, in fact, generalizes Equation (8).

Next, let us consider the map in Equation (1) for a weak invariant. A weak invariant $I(t)$ is a Hermitian operator defined in such a way that its spectrum depends on time, but its expectation value is conserved [5]. It generalizes the Lewis-Riesenfeld invariant [16], which is also a time-dependent Hermitian operator but its spectrum is constant in time (which is referred to in Reference [5] as a strong invariant, in contrast to a weak invariant). An example of a weak invariant is a time-dependent Hamiltonian of a subsystem, the internal energy of which remains constant. Such a concept plays a central role, e.g., in the isoenergetic process in finite-time thermodynamics [17-19]. By definition, $\langle I(t) \rangle = \text{tr}(I(t)\rho(t))$ remains constant in time, that is,

$$\langle I(t') \rangle = \langle I(t) \rangle. \qquad (11)$$

The left-hand side of this equation is calculated to be

$$\langle I(t') \rangle = \text{tr}(I(t')\Phi_{t',t}(\rho(t))) = \text{tr}(\Phi^*_{t',t}(I(t'))\rho(t)), \qquad (12)$$

where $\Phi^*_{t',t}$ stands for the adjoint map of $\Phi_{t',t}$ and is given by

$$\Phi^*_{t',t}(Q) = \sum_n V^\dagger_n(t',t) Q V_n(t',t) \qquad (13)$$

for an arbitrary operator $Q$. From Equations (11)-(13), it follows that $I(t)$ is a weak invariant if it obeys

$$\Phi^*_{t',t}(I(t')) = I(t). \tag{14}$$

This defining equation implies that the adjoint map generates backward time evolution of the weak invariant. Appearance of both the forward and backward dynamical maps is a peculiar aspect of the theory of weak invariants.

Although $I(t)$ has the conserved expectation value by definition, its fluctuation quantified by the variance

$$(\Delta I)^2(t) = \langle I^2(t) \rangle - \langle I(t) \rangle^2 \tag{15}$$

varies, in general. Actually, it does not decrease. This can be seen as follows. Clearly, the second term on the right-hand side in Equation (15) is constant, and so it is sufficient to consider only the second moment

$$\langle I^2(t') \rangle = \mathrm{tr}\left(I^2(t')\rho(t')\right) = \mathrm{tr}\left(\Phi^*_{t',t}(I^2(t'))\rho(t)\right). \tag{16}$$

A crucial point to be noted is that *the adjoint map is not only completely positive itself but also unital,* $\Phi^*_{t',t}(\mathbb{I}) = \mathbb{I}$, *even if* $\Phi_{t',t}$ *is not unital*, as can be seen in Equations (3), (13) and (14). Then, the operator convex $I^2$ satisfies [14]

$$\Phi^*_{t',t}(I^2(t')) \geq \left\{\Phi^*_{t',t}(I(t'))\right\}^2 = I^2(t). \tag{17}$$

Substitution of this inequality into Equation (16) yields $\langle I^2(t') \rangle \geq \langle I^2(t) \rangle$. Therefore,

$$(\Delta I)^2(t') \geq (\Delta I)^2(t) \qquad (18)$$

holds for $t' > t$. This shows how the fluctuation of a weak invariant exhibits a temporally asymmetric behavior.

At the end of this section, a couple of comments are in order. Firstly, the entropies are defined independently of dynamics, whereas a weak invariant is directly associated with the dynamics (i.e., system-specific). Secondly, the entropies show their monotonicity as in Equations (8) and (10) under the assumption that the dynamical map is unital, but the result in Equation (18) does not need such an additional requirement.

## 3. Evolution of Fluctuations and Correlations of Weak Invariants under Gorini-Kossakowski-Lindblad-Sudarshan Equation

Let us consider infinitesimal time evolution in the Markovian approximation. That is, $t'$ in Equation (1) is set to be $t + \Delta t$. Choose one of $V_n$'s, say $V_0$, and write the others $V_i$'s ($i = n \neq 0$). Then, it is convenient to write them as follows [20]:

$$V_0(t + \Delta t, t) = \mathbb{I} - i \Delta t H - \frac{\Delta t}{2} \sum_i |g_i|^2 L_i^\dagger L_i + O((\Delta t)^2), \qquad (19)$$

$$V_i(t + \Delta t, t) = (\Delta t)^{1/2} g_i L_i + O(\Delta t). \qquad (20)$$

Here, $g_i$'s are *c*-number coefficients that can be complex, in general, $H$ is the Hamiltonian of the subsystem *A* and $L_i$'s are the so-called Lindbladian operators. All of them may depend locally on *t*. It is straightforward to ascertain that the trace-preserving condition in Equation (3) is fulfilled by Equations (19) and (20) up to $O((\Delta t)^{3/2})$. Substitution of Equations (19) and (20) into Equation (1) yields

$$\rho(t+\Delta t) = \rho(t) - i\Delta t[H,\rho(t)] - \Delta t \sum_i c_i \left(L_i^\dagger L_i \rho(t) + \rho(t) L_i^\dagger L_i - 2 L_i \rho(t) L_i^\dagger\right) + O\left((\Delta t)^{3/2}\right),$$

where $c_i$'s are nonnegative coefficients given in terms of $g_i$'s in Equations (19) and (20) as follows:

$$c_i = \frac{1}{2}|g_i|^2. \qquad (21)$$

Therefore, in the limit $\Delta t \to 0+$, the Gorini-Kossakowski-Lindblad-Sudarshan equation [6-8]

$$i\frac{\partial \rho}{\partial t} = [H,\rho] - i\sum_i c_i \left(L_i^\dagger L_i \rho + \rho L_i^\dagger L_i - 2 L_i \rho L_i^\dagger\right) \qquad (22)$$

is obtained. This is known to be the most general linear Markovian master equation that preserves positive semidefiniteness of the density matrix. For such preservation, nonnegativity of $c_i$'s is essential. The second term on the right-hand side, which highlights the difference of Equation (22) from the Liouville-von Neumann equation is often called the dissipator.

It is known in the literature how the von Neumann entropy and the Rényi entropy evolve under the Gorini-Kossakowski-Lindblad-Sudarshan equation. The time derivatives of these entropies satisfy the following inequalities [21,22]:

$$\frac{dS[\rho(t)]}{dt} \geq 2 \sum_i c_i \langle [L_i^\dagger, L_i] \rangle, \qquad (23)$$

and

$$\frac{dS_\alpha[\rho(t)]}{dt} \geq 2 \sum_i c_i \langle [L_i^\dagger, L_i] \rangle_\alpha, \qquad (24)$$

respectively. The symbol $\langle Q \rangle_\alpha$ in Equation (24) stands for the so-called $\alpha$-expectation value defined by $\langle Q \rangle_\alpha = \mathrm{tr}(Q \rho^\alpha)/\mathrm{tr}\,\rho^\alpha$. Clearly, Equation (24) becomes Equation (23) in the limit $\alpha \to 1$. Both of these time derivatives are nonnegative if $[L_i^\dagger, L_i] = 0$, that is, the Lindbladian operators are normal [see the discussion just below Equation (4)]. It is seen from Equation (20) that the normal Lindbladian operators correspond to the unital dynamical map $\Phi_{t',t}$ in Equation (1), in consistency with the fact that Equations (8) and (10) hold if the map is unital.

Now, in a similar way, the weak invariant in Equation (14) with Equations (19) and (20) can also be represented in the differential equation of the following form [5,9,23]:

$$i \frac{\partial I}{\partial t} = [H, I] + i \sum_i c_i \left( L_i^\dagger L_i I + I L_i^\dagger L_i - 2 L_i^\dagger I L_i \right). \qquad (25)$$

In fact, the expectation value is conserved

$$\frac{d\langle I(t) \rangle}{dt} = 0 \qquad (26)$$

if Equations (22) and (25) are taken into account.

Now, there exist multiple weak invariants as solutions of Equation (25), in general. In the Appendix, an example is given, in which multiple weak invariants are present. This situation may be in contrast to that concerning the entropies. Let a collection of such weak invariants be denoted by $\{I_K(t)\}_K$. What to be analyzed is the covariance matrix, the $(K, K')$-element of which is given by

$$C(I_K(t), I_{K'}(t)) = \frac{1}{2}\langle\{I_K(t), I_{K'}(t)\}\rangle - \langle I_K(t)\rangle\langle I_{K'}(t)\rangle, \qquad (27)$$

where $\{Q_1, Q_2\} = Q_1 Q_2 + Q_2 Q_1$. From Equations (22) and (25), time evolution of this quantity is calculated to be

$$\frac{d}{dt} C(I_K(t), I_{K'}(t))$$
$$= \sum_i c_i \langle [L_i, I_K(t)]^\dagger [L_i, I_{K'}(t)] + [L_i, I_{K'}(t)]^\dagger [L_i, I_K(t)] \rangle. \qquad (28)$$

A diagonal element $C(I_K(t), I_K(t))$ is the variance $(\Delta I_K(t))^2$ quantifying the fluctuation. Its time evolution is thus nonnegative

$$\frac{d}{dt}(\Delta I_K(t))^2 = 2\sum_i c_i \langle [L_i, I_K(t)]^\dagger [L_i, I_K(t)] \rangle \geq 0, \qquad (29)$$

under the Gorini-Kossakowski-Lindblad-Sudarshan equation. And, in order for this conclusion to be valid, the Lindbladian operators do not have to be normal, corresponding to the fact that Equation (18) holds even for nonunital dynamics maps. This is in marked contrast to the case of the entropies. On the other hand, the off-diagonal elements of the covariance matrix do not have any such temporally asymmetric property, suggesting that nontrivial information about the subdynamics may be contained in the off-diagonal elements. This point is yet to be clarified. It is also mentioned that Equations (23), (24) and (29) show that the temporally monotonic behaviors are certainly connected to the dissipator, as they should be.

## 4. Concluding Remarks

In the present work, it has been discussed how the fluctuations of weak invariants associated with the quantum subdynamics of an open quantum system exhibit temporally monotonic behavior. This result has been considered in comparison with that of the entropy, specifically the Rényi entropy as well as the von Neumann entropy. The entropies do not decrease in time if a completely positive map describing the subdynamics is unital, whereas the fluctuations of the weak invariants monotonically grow without the unital condition on the map: the trace-preserving condition in Equation (3) is sufficient. A formula has also been derived for time evolution of the covariance matrix of the weak invariants under the Gorini-Kossakowski-Lindblad-Sudarshan equation.

After completion of this work, the present authors' attention has been drawn to earlier ones in References [24,25]. There, the damped quantum harmonic oscillator is treated by use of the Gorini-Kossakowski-Lindblad-Sudarshan equation, and time evolution of the linear entropy, $S^{(L)}[\rho] = 1 - \text{tr}\,\rho^2$, as well as the von Neumann entropy in such a specific system is discussed. Since the Rényi entropy $S_\alpha[\rho]$ in Eq. (9) with $\alpha = 2$ is related to the linear entropy as $S^{(L)}[\rho] = 1 - \exp(-S_{\alpha=2}[\rho])$, the results in Equations (10) and (24) are more general than those in References [24,25].

Now, the comparison between the weak invariants and entropies in the context of temporal asymmetry may shed light on a possible hidden relationship between them. In a recent work [26], the action principle has been formulated for a quantum master equation based on the auxiliary field formalism [27], and it has been found that the auxiliary field is actually a weak invariant. In addition, the auxiliary field has also shown to be the Noether charge [28]. This reminds one of another observation that the black hole entropy related to surface gravity can be regarded as the Noether charge [29]. Although so far these two facts are apparently independent each other, the underlying doubled structures, "auxiliary field and density matrix" and physical quantities "inside and outside" the Killing horizon, seem to indicate a more general feature to be explored, apart from the (non)relativistic nature. Regarding this point, it should be noticed that the authors of Reference [27] mention an analogy between the doubled structure of "auxiliary field and density matrix" and the tilde conjugation symmetry in thermo field dynamics [30]. In addition, the authors of Reference [31] make a comment that the doubled structure in thermo field dynamics might have its origin in the spacetime structure, although it remains speculative.

**Appendix**

It may be useful to give an explicit example of the Gorini-Kossakowski-Sudarshan equation that admits multiple weak invariants. The subsystem considered here is the time-dependent harmonic oscillator with unit mass. In this case, the equation may physically describe a one-dimensional analog of the subdynamics of the time-dependent harmonic trap of an atom surrounded by the dissipative environment, for example. The Hamiltonian reads $H(t) = p^2/2 + k(t)x^2/2$, where $k(t)$ is a nonnegative time-dependent *c*-number coefficient corresponding to the frequency squared. In Reference [17], it is shown that if this Hamiltonian is required to be a weak invariant, then the dissipator term in the Gorini-Kossakowski-Lindblad-Sudarshan equation is determined without detailed knowledge about the interaction between the subsystem and the environment. The discussion can drastically be simplified by introduction of the following operators: $K_1 = p^2/2$, $K_2 = x^2/2$ and $K_3 = (px + xp)/2$. These satisfy the commutation relations: $[K_1, K_2] = -iK_3$, $[K_2, K_3] = 2iK_2$ and $[K_3, K_1] = 2iK_1$, which implies that the operators form the Lie algebra isomorphic to *su*(1,1). $K_3$ is the generator of the dilatation transformation that plays a central role in realizing the squeezed state [32]. In terms of these operators, the Hamiltonian is expressed as

$$H(t) = K_1 + k(t)K_2. \tag{A1}$$

In order for this Hamiltonian to be a weak invariant of the Gorini-Kossakowski-Lindblad-Sudarshan equation, it should fulfill Equation (25) as *I*. Such an equation together with the Lie-algebraic structure can put a stringent condition on the form of the dissipator. Accordingly, the Gorini-Kossakowski-Lindblad-Sudarshan equation is found to be given by [17]

$$i\frac{\partial \rho}{\partial t} = [H, \rho] + \frac{i}{2}\dot{k}\left[K_2, [K_2, \rho]\right], \quad (A2)$$

provided that $\dot{k} \equiv dk(t)/dt < 0$ should hold because of the condition in Equation (21). Correspondingly, Equation (25) becomes

$$i\frac{\partial I}{\partial t} = [H, I] - \frac{i}{2}\dot{k}\left[K_2, [K_2, I]\right]. \quad (A3)$$

Clearly, the Hamiltonian in Equation (A1) is a solution of this equation. Write the Hamiltonian as

$$I_1(t) \equiv H(t). \quad (A4)$$

Now, another weak invariant may be expressed as a linear combination of $K_i$'s:

$$I_2(t) = \alpha_1(t)K_1 + \alpha_2(t)K_2 + \alpha_3(t)K_3, \quad (A5)$$

where $\alpha_i(t)$'s are $c$-number coefficients. $[I_1(t), I_2(t)] \neq 0$ if $\alpha_2(t) \neq k(t)\alpha_1(t)$ and $\alpha_3(t) \neq 0$. This quantity is a solution of Equation (A3) if the coefficients satisfy

$$\frac{d}{dt}\begin{pmatrix} \alpha_1(t) \\ \alpha_2(t) \\ \alpha_3(t) \end{pmatrix} = A(t) \begin{pmatrix} \alpha_1(t) \\ \alpha_2(t) \\ \alpha_3(t) \end{pmatrix} \quad (A6)$$

where $A(t)$ is a matrix given by

$$A(t) = \begin{pmatrix} 0 & 0 & -2 \\ \dot{k}(t) & 0 & 2k(t) \\ k(t) & -1 & 0 \end{pmatrix}. \tag{A7}$$

The solution of this equation is

$$\begin{pmatrix} \alpha_1(t) \\ \alpha_2(t) \\ \alpha_3(t) \end{pmatrix} = U(t) \begin{pmatrix} \alpha_1(0) \\ \alpha_2(0) \\ \alpha_3(0) \end{pmatrix}, \tag{A8}$$

$$U(t) = \mathcal{T} \exp\left[\int_0^t dt'\, A(t')\right], \tag{A9}$$

where $\mathcal{T}$ denotes the chronological symbol.

**Author Contributions:** S.A. has provided the idea of research. Both of the authors have performed the calculations, written the manuscript, and approved the final version of the manuscript.

**Funding:** The work of S.A. has been supported in part by a grant from National Natural Science Foundation of China (No. 11775084), the Program of Fujian Province and the Program of Competitive Growth of Kazan Federal University from the Ministry of Education and Science of the Russian Federation.

**Conflicts of Interest:** The authors declare no conflict of interest.